\input harvmac 
\baselineskip=12pt
\overfullrule=0pt

\def\bgamma{(\bar\beta, \beta)}
\def\Tr{{\rm Tr~}}
\def\today{\ifcase\month\or
  January\or February\or March\or April\or May\or June\or
  July\or August\or September\or October\or November\or December\fi
  \space\number\day, \number\year}
\hskip4in{IC/96/273}


\hskip4in{hep-th/9612172}
\vskip1in
\centerline{\bf Relating 
$c<0$ and $c>0$ Conformal Field Theories} \vskip1cm 
\centerline{Sathya Guruswamy$^1$ and Andreas W. W. Ludwig$^{2}$}
\vskip0.1in
\centerline{\it ${}^1$International Center for Theoretical Physics}
\centerline{\it P.O. Box 586, Trieste 34100 ITALY} 
\vskip0.1in
\centerline{\it and} 
\vskip0.1in
\centerline{\it ${}^2$Department of Physics, University of California} 
\centerline{\it Santa Barbara, CA-93106, USA}
\vskip0.5in
\vskip0.1in
\lref\mudry{ C. Mudry, C. Chamon and X. G. Wen, Phys. Rev. B53 (1996) 
R7638; Nucl. Phys. B466 (1996) 383.}
\lref\bernard{ D. Bernard, Cargese Lectures;  hep-th/9509137.}
\lref\maass{ Z. Maassarani and D. Serban, hep-th/9605062.} 
\lref\readposdef{For a discussion as to how this is achieved
within the framework of the respective $c<0$ theories,
see \readmilan. }
\lref\IQHEfermion{The description of the Integer Quantum Hall
transition using $c=-1$ bosonic ghosts (and graded supersymmetry)
is based on \LFSG \ and \Lee. }
\lref\motivBosonZero{The $c=2$ character arising this
way should be viewed here, as mentioned above, as the product
of the  character
of a complex scalar at zero boson radius, and a single
decoupled oscillator zero mode.}
\lref\suOneOne{See for example: N. Ja. Vilenkin, A. U. Klimyk,
``Representation of Lie Groups and Special Functions'',
(Kluwer Academic Publ., Boston, 1993).}
\lref\FMS{A comprehensive discussion is given in: D. Friedan, E. Martinec 
and S. Shenker, Nucl. Phys. B271 (1986) 93.}
\lref\wenwu{X. G. Wen, Y. S. Wu, T. Hatsugai, Nucl. Phys. B422 (1994) 476;
X. G. Wen, Y. S. Wu, Nucl. Phys. B419 (1994) 455.}
\lref\Goddard{P. Goddard, D. Olive and G. Watterson, Commun. Math. Phys. 
112 (1987) 591. } 
\lref\commentLeClair{The  ${\hat sl}(2)$ symmetry
discussed in \leclair \ appears to be local, and
thus different from ours.}
\lref\GLrandom{S. Guruswamy and A. W.W. Ludwig, {\it in preparation}.}
\lref\saleurpolymer{  H. Saleur, Nucl. Phys. B382 (1992) 486.   } 
\lref\yang{C. N. Yang, Phys Rev Lett (1989) 2144.}
\lref\kausch{H. Kausch, hep-th/9510149.} 
\lref\readmoore{G. Moore and N. Read, Nucl. Phys. B360 (1991) 362.} 
\lref\readmilan{M. Milovanovic and N. Read, Phys. Rev. B53 (1996) 13559.} 
\lref\baake{M. Baake, J. Math. Phys.29 (1988) 1753.} 
\lref\Ginsparg{P. Ginsparg, Les Houches.} 
\lref\kacwakimoto{V. G. Kac and M. Wakimoto, Proc. Nat. Acad. Sci. 85 
(1988) 4956.}
\lref\pasquier{D. Bernard, V. Pasquier, D. Serban, 
Nucl. Phys. B428 (1994) 612.}
\lref\haldane{F. D. M. Haldane et al, Phys Rev Lett 69 (1992) 2021.}
\lref\schout{ P. Bouwknegt, A. W.W. Ludwig, K. Schoutens,
 Phys Lett B338(1994) 448.}
\lref\leclair{A. LeClair,  Nucl. Phys. B415 (1994) 734.}
\lref\BFZ{L. Balents, M. P.A. Fisher and M. Zirnbauer, cond-mat/9608049.}
\lref\AffLudwig{I. Affleck and A. W.W. Ludwig, Nucl. Phys. B360 (1991) 641.}
\lref\FLS{P. Fendley, A. W.W. Ludwig and H. Saleur, Phys. Rev. Lett. 74 
(1995) 3005.}
\lref\efetov{ K. B. Efetov, Adv. in Physics 32 (1983) 53.}
\lref\readnew{ N. Read and I. Rezayi, cond-mat/9609079.}
\lref\commentreadsuOneOne{ See also \readnew.}
\lref\schoutthanks{We thank K. J. Schoutens for clarifying discussions
on this point.}

\vskip1in
\centerline{\bf ABSTRACT}
\vskip0.1in

A `canonical mapping'  is established
  between the $c=-1$ system of bosonic ghosts 
 and the $c=2$ 
complex scalar theory {\it and}, a similar mapping between the $c=-2$ 
system of fermionic ghosts  and the $c=1$ Dirac 
theory. 
The existence of this mapping is suggested  by the identity of the 
characters of the respective theories.
The respective $c<0$ and $c>0$ theories share the same space
of states, whereas the spaces of conformal
fields are different.
Upon this mapping from their $c<0$ counterparts,   
the ($c>0$)  complex scalar and the Dirac theories inherit
 hidden nonlocal   $sl(2)$ symmetries.

\vfill
\eject

\newsec{Introduction}
\vskip0.1in

Two dimensional conformal
field theories  with
central charge  $c<0$ have recently attracted much
attention, especially in the context
of condensed matter applications such as 
unusual (paired) Quantum Hall states \refs{\readmoore,
\wenwu, \readmilan}, 
disordered systems \refs{\mudry,\bernard,\maass},
including the plateau transition
\nref\LFSG{A. W.W. Ludwig, M. P.A. Fisher, R. Shankar and G. Grinstein,
Phys. Rev. B50 (1994) 7526.}%
\nref\Lee{ D.H.Lee, Phys Rev B50 (1994) 10788.}%
in the Integer Quantum Hall effect \IQHEfermion \  
and  polymer theories \saleurpolymer.
All of these systems involve the 
 so called $(\beta, \gamma)$ 
system of conformal weight $(1/2, 1/2)$ ``bosonic ghost'',
of central charge $c=-1$, or  a version of the
fermionic ghost $(\xi,\eta)$ 
system  of central charge $c=-2$.
 (Both these systems were 
first introduced in
the context of string theories \FMS).

Both are non-unitary conformal field theories,
the space of states created by the $c<0$ Virasoro
algebra containing negative norm states.
This is a complication for various reasons.
For example, in the context of the mentioned Quantum Hall systems,
the spectrum of the physical Hamiltonian of the edge states
of the system is given in terms of the Virasoro zero mode
$L_0$.
 Clearly, the Hilbert space of states 
on which it acts must be positive definite \readposdef.
In the context of disordered systems,
the bosonic ghost system interacts with a Dirac fermion.
Certain  exactly
marginal interactions can be a treated exactly \refs{\mudry,
\LFSG}.
 However, in general the system flows off to strong coupling.
Very little intuition is available to help
access the strong coupling behavior,
especially because of the non-unitary
nature of the bosonic ghost part of the theory.  
Therefore, a better understanding,
and a better intuition about the properties
of the $c<0$ ghost systems is highly desirable.

In this paper
we show that  there is a simple
relationship  (i) between  the
fermionic ghost theory
at $c=-2$ and the free Dirac theory at $c=1$,
as well as  (ii) between the
$c=-1$  bosonic ghost theory and
a $c=2$ theory of a complex scalar field.
First, we observe
that the characters of the respective theories,
i.e.  (i) of $c=-1$ and $c=+2$
as well of (ii) of $c=-2$ and of $c=+1$,
are identical, in the various sectors
[periodic and antiperiodic]. 
This suggests a deeper connection
between the respective theories.
Indeed there is. We establish a `canonical'
mapping from one theory to the other.
This mapping  implies that the {\it space of states} of
the $c<0$ theories is identical to that
 of the respective  $c>0$ theories, and so are the  
{\it $L_0$-generators} (upto a shift in ground state energy due to 
different normal ordering prescriptions).  Therefore,
we have two 
different conformal field theories 
which share the same spectrum of $(L_0-c/24)$
and the same space of states  (implying that the characters
are the same).
 The spaces of conformal fields
 of the respective $c<0$ and $c>0$ theories
 are different, and the fields 
 are  nonlocally related to each other.
The Virasoro generators $L_n$ with $n \not = 0$ are not
simply related upon this mapping.

This mapping allows us to 
 establish the existence of {\it hidden non-abelian symmetries}
in the theories with $c>0$.
These are not visible in the usual way from the Lagrangian,
since there is no corresponding Noether current.
The bosonic ghost theory is known
to possess an  $ sl(2)$  Kac-Moody current algebra
 symmetry  at level $k=-1/2$. 
  Our mapping establishes a $sl(2)$ current
algebra symmetry
also for the $c=2$ theory of a complex scalar field,
for half-integer moding.
A complex scalar theory has an obvious $U(1)$ symmetry,
and an associated global charge.
However, its Noether current does not 
 give rise to a Kac-Moody
current algebra symmetry, because of logarithms
in its correlation functions. But  our mapping shows
that the global $U(1)$ charge  is nevertheless
the zero mode of a  {\it hidden} $U(1)$ KM current
algebra.
Moreover, we show that there are two
additional  sets of current algebra generators,
 that extend this $U(1)$ 
current algebra symmetry to
an $sl(2)$ current algebra.
Already the global $sl(2)$ symmetry, present
for integer and half-integer moding,  is interesting.
No such symmetry  can be identified from
the Lagrangian of the $c=2$ theory of a complex scalar field. 
The two additional charges, besides the
 $U(1)$ charge, are {\it non-local} when
expressed in terms of the complex scalar field
variables.
Physically, these two generators effect
continuous rotations of the complex scalar
field $\phi(x)$ 
into its charge conjugate  $\phi^{\dagger}(x)$ .
In the Hilbert space of states,  the symmetry
corresponds to
 global rotations between positive and
negative frequency modes.
In addition, we observe that the space of states of the
 the bosonic ghost and
complex scalar  field theories 
has a structure
isomorphic to the multiplets
of the  $SU(2)$-level-1
{\it  Yangian}
\refs{\haldane,\schout, \pasquier}.

The fermionic theory at $c=-2$,
 on the other hand,
is known \refs{\readmilan, \kausch},
 to possess a  global $sl(2)$ symmetry.
 No non-abelian symmetry can be identified from the Lagrangian
for the Dirac fermion. 
Using our mapping, 
the theory of a Dirac fermion (at $c=1$)
inherits a {\it hidden $sl(2)$} symmetry
from the $c=-2$ theory.
 The origin of this symmetry
is the charge conjugation (``particle-hole'')
 symmetry of the Dirac theory.
One of the non-abelian  
global charges is the ordinary Dirac charge.
The remaining two non-abelian charges
are {\it non-local} in the Dirac field
variables. Physically, those two generators
effect continuous rotations of the
Dirac field $\psi(x)$ 
into its charge conjugate $\psi^{\dagger}(x)$. 
(This may be viewed as  a one-species version of the hidden $SU(2)$
symmetry, discovered in the Hubbard model \yang. 
In that model this symmetry is local.)

\vskip0.1in
The organization of this paper is as follows: In 
section 2,
we  define the theories in a language that
will be useful for our discussion
and  exhibit their respective characters.
We see that the characters
of the $c<0$  bosonic and fermionic
theories are identical to  those
of the corresponding $c>0$ bosonic
and fermionic theories for periodic
and antiperiodic 
boundary conditions on the torus.
In section 3,
we give the exact relationship between
 these theories using a  `canonical mapping',
showing that the spaces of states of these theories
are identical.
In section 4, 
we discuss the hidden
non-abelian $sl(2)$
 global and current algebra
 symmetries  that the $c>0$ theories, Dirac and the complex scalar,
inherit from this mapping and we derive the
corresponding non-local generators. 
 Some concluding remarks are made in  section 5. 
(In Appendix A we summarize the results for the characters in the Ramond 
and Neveu-Schwarz sectors for the $c<0$ and $c>0$ bosonic and fermionic 
theories in question, and in Appendix B we discuss the relationship
of the bosonic ghost characters with those derived by 
Kac and Wakimoto for the  $sl(2)_{-1/2}$ current algebra.)

\newsec{ Conformal field theories at $c=\pm 1$ and $c=\pm2$}

\vskip .2cm

\noindent{\it Dirac Fermion ($c=+1$)}:
\vskip0.1in
Let us start
considering a free
 Dirac fermion
described by the action
$$
S_{\psi}=\int d\tau \int dx 
\{
\psi_L^{\dagger}[\partial_{\tau} + i \partial_x]
 \psi_L
+
\psi_R^{\dagger}[\partial_{\tau} - i \partial_x]
 \psi_R
\}
$$
on a torus of size $\beta$ in the ``euclidean time'' direction $\tau$
and of size $l$ in the ``space'' direction $x$.
We will only consider the, say, L-moving part, and drop
the
subscript ${}_L$, i.e. we write  $\psi_L \to \psi$.
We will do this for all theories considered in this paper.
For the chiral field,we  consider a general twisted
boundary condition in the functional integral:
$$
\psi(\tau, x+l) =(-1) e^{2\pi i\lambda'} 
\psi(\tau, x),
\qquad
\psi(\tau+\beta , x) =(-1) e^{ 2\pi i \mu} 
\psi(\tau, x). 
$$
Note that when $\mu=\lambda'=0$ we have antiperiodic
b.c.'s along the spatial and temporal cycles,
``natural'' for a fermionic theory.
The hamiltonian 
can be expressed
in terms of momentum modes
of the fermion operators 
$$
\psi(\tau,x)
= \sum_{ s \in  Z +{1\over 2} -\lambda'}
e^{-(\tau+i x) 2 \pi s /l}
 \ \psi_{s},
\quad
\psi^{\dagger}(\tau,x)
= \sum_{ s \in  Z +{1\over 2} -\lambda'}
e^{(\tau+i x) 2 \pi s /l}
 \ \psi^{\dagger}_{s},
$$
where $p={2\pi\over l} s$
are the momenta of the system
on a space of size $l$:
$$
H_{\psi}
= ({2 \pi/l})
\sum_{s}
s  \psi^{\dagger}_s \psi_s .
$$
We write the general
Dirac character as

\eqn\diracphase{
\chi_{c=+1}(q;\mu ;\lambda')=
\Tr e^{-\beta H_{\psi}}=
q^{-1/24} \ q^{({\lambda'}^2)/2} \ 
\prod_{n=0}^{\infty}
(1+ e^{2 \pi i \mu}
 q^{n+1/2-\lambda' })
(1+  e^{-2 \pi i \mu}
 q^{n+1/2+\lambda' })
}
where the trace is taken over the Hilbert space
of the  fermion modes ( the L-moving part of the theory), acting
on the Fermi sea vacuum 
[giving rise to the prefactor $q^{-1/24}$].
We have set
$ q=e^{-2 \pi \beta/l}$
and $ e^{-2 \pi i \mu}$
is a ``fugacity'' which keeps track of the $U(1)$ charge
of the fermion state (with respect
to filled Fermi sea).
The parameter $\lambda'$ plays the role of
a simple ``phase shift''
of the Dirac fermion,
and shifts the single particle
energy levels
occupied by the fermions
by an amount $ (2\pi/l) \lambda'$.
 In particular, it allows
us to continuously interpolate between antiperiodic
and periodic boundary conditions in the spatial direction.
The prefactor $ q^{({\lambda'}^2)/2} $ 
arises from the shift of the ground state energy
of the filled Fermi sea. 
For $\lambda'=1/2$,
one has periodic spatial  boundary conditions
corresponding  to the Ramond sector.
{}From the character in \diracphase\ we
see that, in this case,
the vacuum is two fold degenerate,
the two ground states having $U(1)$ charges
$Q=0,+1$, as seen from the powers of the fugacities.
The zero modes appear since
for $\lambda'=1/2$ 
there is a single particle state sitting right at
the Fermi level, and therefore,  occupying it does not cost
any energy. 
The vacuum with periodic boundary conditions (the Ramond vacuum)
has conformal weight
 which is an amount of $\Delta=1/8$
above the conformal weight $=0$ of the ground
state with antiperiodic boundary conditions
(Neveu-Schwarz vacuum).

\vskip .5cm
\noindent{\it Bosonic ghost ($c=-1$)}:
\vskip0.1in
The graded supersymmetry approach to  
random
systems  of recent interest
is based on the 
observation
that the Dirac partition function of
\diracphase\
(corresponding to a first derivative action
and hamiltonian)
is exactly cancelled
by a bosonic counterpart.
What is needed for this cancellation
to occur is a theory of bosons with first order
action  having the same boundary conditions
along both cycles of the torus as the fermionic
counterpart. For $\mu=0$ this implies 
  {\it anti}-periodic b.c.'s
along the time direction,
`unnatural' for bosons.
A suitable bosonic action is that of
the bosonic ghost 
 system (known as the ``beta-gamma'' system
in string theory\FMS).
Since this cancellation occurs for
arbitrary inhomogeneous quadratic 
potentials (``random potentials''),
it can be used to represent averages over those
random potentials in terms of
a theory of interacting Dirac fermions
and bosonic ghosts \refs{\efetov,\mudry}.

The first order action (for the L-moving part)
 is given \Goddard\ 
in terms
of two real variables $\beta_1(\tau,x),\beta_2(\tau,x)$:
$$
S_{\beta}
=
{-i\over 2}
\int d \tau \int d x
\epsilon^{ab}
\beta_a
[\partial_{\tau} + i \partial_x]
\beta_b
$$
$\epsilon^{ab}$,
$\epsilon^{12}=-\epsilon^{21}=1$,
 is the antisymmetric tensor.
( Often the notation $\beta_1 \to \gamma$ and
$ i \beta_2 \to \beta$ is used \FMS).
The action is invariant
under the symplectic group 
$Sp(2)$,
under which the fields transform as  a doublet.
We define the linear combinations
$$
\beta ={1\over \sqrt{2}} [ \beta_1 + i \beta_2],
\qquad {\bar \beta} ={1\over \sqrt{2}} [ \beta_1 - i \beta_2].
$$
The hamiltonian, as obtained from
the canonical stress tensor, can
be written in terms
of momentum
modes, 
$$
\beta(\tau,x)
= \sum_{ s \in  Z +{1\over 2} -\lambda'}
e^{-(\tau+i x) 2 \pi s /l}
 \ \beta_{s},
\quad
{\bar \beta}(\tau,x)=
 \sum_{ s \in  Z +{1\over 2} -\lambda'}
e^{(\tau+i x) 2 \pi s /l}
 \ {\bar \beta}_{s}
$$
satisfying canonical commutations relations
$[{\beta}_s, {\bar \beta}_t]=\delta_{s,t}$,
as:
$$
H_{\beta}
= ({2 \pi/l})
\sum_{s}
s  {\bar \beta}_s \beta_s .
$$
Since  the momentum mode index  $s$ can take on negative values, 
the hamiltonian is normal ordered on a vacuum
satisfying 
$\beta_s|0> = {\bar \beta}_{-s} |0>=0$
for $ s \geq 0$ (operators
annihilating this vacuum are moved to the right).
The normal ordered hamiltonian is bounded from below.
[In the context of disordered systems,
this choice of ghost vacuum is dictated
by the convergence of the bosonic
part of the functional integral that
cancels the corresponding fermionic one.]
Since the partition function factorizes
it
is sufficient to consider a single momentum $p
= {2 \pi \over l} s$.
Consider
the bosonic partition
function for a single mode (excluding the ground state energy)
$$
Z_{\beta_s}
 \equiv 
\Tr~ 
e^{-[\beta  \  (2 \pi /l) s - 2 \pi i ( {1\over 2} + \mu)]
  \ :{\bar \beta}_s\beta_s :}
$$
where the  trace is taken over the bosonic Fock
space of the canonical operators.
Note that for $\mu=0$ we have inserted
$(-1)^N$ where $N$ is the boson number.
This insertion corresponds to antiperiodic b.c.'s
in euclidean time $\tau$ in the functional
integral formulation,
in which the bosonic fields satisfy
the general twisted b.c.'s:
$$
\beta(\tau, x+l) =(-1) e^{2\pi i\lambda'} 
\beta(\tau, x), 
\qquad
\beta(\tau+{1\over T}, x) =(-1) e^{ 2\pi i \mu} 
\beta(\tau, x) 
$$
where we denote the inverse temperature
by $\beta = 1/T$ [ No confusion should arise between
the inverse temperature and bosonic ghost field.]
This yields 
the following chiral  partition function 
\eqn\betagammaAAla{
\chi_{c=-1} (q,\mu,\lambda')
=
q^{-(-1/24)} \ q^{-{1\over 8} +({1\over 2} - {\lambda'})^2/2} \ 
\prod_{n=0}^{\infty}
{1\over (1+ e^{2\pi i \mu} q^{n+1/2-\lambda'})}
 \
{1 \over (1+ e^{-2\pi i \mu} q^{n+1/2+\lambda'})}.
}
[The prefactor of
$
q^{-(-1/24)} \ q^{-{1\over 8} +({1\over 2} - {\lambda'})^2/2}  $
 arises from normal ordering on the ghost vacuum.]
For half-integer moding ($\lambda' =0$),
the combined theory of a chiral Dirac and
a bosonic ghost field, relevant in the context of
disordered systems, possesses graded supersymmetry,
reflected in the fact that the product of the
two partition functions is identically  unity
on the torus (Appendix A).
The case of integer moding, represented e.g. by
a phase shift of $\lambda' =1/2$
requires extra discussion.
In this case,  the cancellation
of the fermionic partition function by the bosonic one
is achieved by  an extra  single bosonic zero mode
$\beta_0$  (annihilating the vacuum).
The part of the  partition function coming
from  the bosonic zero  mode
needs to be regularized:
$$
Z_{\beta_0} =
{1 \over (1+ e^{2\pi i \mu} )}=
\lim_{\eta\to 0^+}
\sum_{n=0}^{\infty}
(-1)^n e^{n(2\pi i \mu - \beta \eta )}.
$$
This cancels precisely the contribution
from the Dirac zero mode,
$\psi_0$,
which is
$$ Z_{\psi_0}
=(1+ e^{2\pi i \mu} ).
$$ 
Note that the expansion
in powers of the fugacity shows
that
the bosonic ghost ground state
has an infinite degeneracy, arising
from the zero mode.
[The zero-mode Hilbert space
is  to be thought of as a sum of  the two irreducible
lowest weight representations
of the non-compact group $SU(1,1)$  of a single boson
(see Appendix B for a brief discussion)].
\vskip .5cm
\noindent{\it Complex scalar field ($c=+2$)}: 
\vskip0.1in
Consider a theory 
of two real scalar fields, $\varphi_j$  $(j=1,2$),
or equivalently, a complex scalar field, $\phi^{\dagger},\phi$,
with action
\eqn\scalaraction{
S =
{1\over 8 \pi}
\sum_{j=1,2}
\int d \tau \int dx 
(\partial \varphi_j)^2=
{1\over 8 \pi}
\int d \tau \int dx 
(\partial \phi^*)
(\partial \phi)
}
where 
$$
\phi(\tau,x)
= {1\over \sqrt{2}}
[
\varphi_1(\tau,x)
+ i
 \varphi_2(\tau,x)
],
\qquad
\phi^{\dagger}(\tau,x)
= {1\over \sqrt{2}}
[
\varphi_1(\tau,x)
- i
 \varphi_2(\tau,x)
].
$$

As usual, from the equations of motion,
 $\phi = \phi_L(z) + \phi_R(z^*)$
and we focus on $\phi_L \to \phi$ (dropping subscript ${}_L$).
First, consider antiperiodic boundary conditions in
space. There is no zero mode
and
the usual momentum space mode expansion
reads
$$
\varphi_j(\tau, x)
=
 \sum_{ s \in  Z +{1\over 2}, s>0}
{1\over \sqrt{s} }
\{
b_{j,s}
e^{-(\tau+i x) (2 \pi/l) s }
+
b^{\dagger}_{j,s}
e^{+(\tau+i x) (2 \pi/l) s }
\},
\qquad (j=1,2)
$$
where 
$$
[b_{j_1,s_2},b^{\dagger}_{j_2,s_2}] =\delta_{j_1,j_2}
\delta_{s_1,s_2}, \qquad (s_1,s_2 > 0)
$$
are oscillator modes,
quantized on the usual  vacuum 
$b_{j,s}|0>=0$,
 and occupying single
particle levels  with positive momenta
$p ={2 \pi \over l} ( n + 1/2) $, $n=0,1,2,3,..$.

So far we have  used antiperiodic
b.c.'s. However, for the complex scalar
we may define more general twisted boundary
conditions
$$
\phi(\tau,x+l)
=(-1) e^{2 \pi i\lambda'} \phi(\tau,x).
$$
When $-1/2 < \lambda' < 1/2$,
this implies the following mode expansion of the
complex scalar field.
In the formula below, the mode indices
of the canonical operators
take on values $s\in Z + 1/2 -\lambda'$
  for $B_{+,s}$ and 
values $t\in Z + 1/2 +\lambda'$
for $B_{-,t}$: 
\eqn\complscalarmodes{
\phi(\tau,x)
 =
\sum_{ s >0}
{B_{+,s}\over \sqrt{s}}
e^{-(\tau+i x) (2 \pi/l) s }
+
\sum_{t>0}
{B^{\dagger}_{-,t} \over \sqrt{t}}
e^{+(\tau+i x) (2 \pi/l) t }
}
and

\eqn\complscalarmodesstar{
\phi^{\dagger}(\tau,x)
=
 \sum_{ t>0}
{ B_{-,t}\over \sqrt{t}}
e^{-(\tau+i x) (2 \pi/l) t }
+
 \sum_{  s>0}
{B^{\dagger}_{+,s}\over \sqrt{s}}
e^{+(\tau+i x) (2 \pi/l) s }
}
($ B_{\pm,s} =$
$ [b_{1,s} \pm i b_{2,s} ]/\sqrt{2},$
when $\lambda'=0$.)
The hamiltonian is

\eqn\scalarhamiltonian{
H =
 {2\pi \over l}
\Bigl \{ 
 \sum_{ s>0}
 \
s  \
[ 
B^{\dagger}_{+,s}B_{+,s} +1/2]
+
 \sum_{ t>0}
t \  [B^{\dagger}_{-,t}B_{-,t}+1/2]
\Bigr \}.}
The partition function factorizes so that for each mode,
say $B_{+,s}$
we define (excluding the ground state energy)
$$
Z_{B_{+,s}}
\equiv
\Tr~ 
\exp \{
-[\beta {2\pi\over l} s- 2 \pi i ( {1\over 2 }+ \mu)]
B^{\dagger}_{+,s}B_{+,s} 
 \}
$$
which gives the
character for the
complex scalar

\eqn\complexSAAla{
\chi_{c=+2}(q,\mu,\lambda')
=
q^{-(2/24)} \ q^{{1 \over 8} - ({\lambda'})^2/2}
 \ 
\prod_{n=0}^{\infty}
{1\over (1+ e^{2\pi i \mu} q^{n+1/2-\lambda'})}
 \
{1 \over (1+ e^{-2\pi i \mu} q^{n+1/2+\lambda'})},
}
[The prefactor of 
$q^{-(2/24)} \ q^{{1 \over 8} - ({\lambda'})^2/2}$
arises from the ground state energy of the
oscillators.] 
We see that this
is the character of the bosonic ghost,
up to an overall power of $q$.
As we move $\lambda'$ from $0$ to $\lambda'=1/2$ a single
 bosonic zero mode
$B_{+,0}$
occurs. Thus,  for integer moding ($\lambda'=1/2$),
 the character
is a product of that of  a complex scalar field, at zero boson
radius, and a decoupled single oscillator zero mode.

\vskip.5in

\noindent{\it Fermionic ghost and Symplectic fermion ($c=-2$)}:
\vskip0.1in
Finally we consider the fermionic analog
of the complex scalar theory, known
 also as the  ``symplectic fermion'' 
for reasons that will be obvious shortly.
This theory is related to the conformal weight $(0,1)$  
fermionic ghost system $(\xi, \eta)$  with $c=-2$ 
system in string theory.
We follow here Ref. \kausch, where
details relating the $(\xi, \eta)$ 
system to the symplectic fermions can also  be found.
[A  similar analysis of this theory can be found in \readmilan.]
 The symplectic 
fermions are described by a Grassman functional integral over two fermion 
fields, ${\bf \Phi}^{a}$, $a=\pm$,
 based on the second derivative action
$$
S= {1\over 4 \pi}
\int d \tau \int d x
 \epsilon_{ab}
  \partial_{\mu} {\bf \Phi}^{a}
  \partial^{\mu} {\bf \Phi}^{b}
$$
where  $\epsilon^{+-}=-\epsilon^{-+}$,
 $\epsilon_{ab}=-\epsilon^{ab}$.
The action is invariant under symplectic
transformations, the field
transforming  as a doublet.
Just as for the bosonic scalar field,
the general solution of the equation
of motion
shows that the field
is a sum of analytic and anti-analytic parts
($z=\tau + i x$):
$$
{\bf \Phi}^{a}(\tau,x)
=
\Phi_L^{a}(z)
+
\Phi^{a}_R(z^*). 
$$
Invariance of the action
under shift by a constant Grassman number,
$\delta \Phi^a =  \theta^a$,
results in
(anti-) analytic conserved fermionic currents
$$ 
{\cal J}_{ L}^a (z) \equiv i \partial_z {\bf \Phi}^a,
\qquad
{\cal J}_{R}^a (z^*) \equiv i \partial_{z^*}{\bf  \Phi}^a. 
$$
{}From the action one obtains the canonical
stress tensor,
$$
T_L(z) = {-1\over 2} \epsilon_{ab}
(\partial_z {\bf \Phi}^a)
(\partial_z {\bf \Phi}^b),
\qquad
T_R(z^*) = {-1\over 2} \epsilon_{ab}
(\partial_{z^*} {\bf \Phi}^a)
(\partial_{z^*} {\bf  \Phi}^b)
$$
which yields the hamiltonian
$$
H = H_L + H_R
$$
where
\eqn\hamsympl{
H_L
=
{-1\over 2}
\int d x
[
(\partial_x \Phi^+_L)
(\partial_x \Phi^-_L)
-
(\partial_x \Phi^-_L)
(\partial_x \Phi^+_L)
]
}
(and a  similar expression for the R-moving part).

First consider a field $\Phi^a_L$
which is antiperiodic
on finite space (size $l$).
 Using an expansion in Fourier modes,
and dropping the subscript ${}_L$
\eqn\fermioncurrents{
\Phi^{\pm}(\tau,x)
= \sum_{s \in Z + 1/2}
{
{\cal J}_{s}^{\pm}\over s}
e^{ {2 \pi \over l} i s (\tau + i x)}
}
(due to antiperiodicity the zero mode is absent)
where
the ``fermionic currents''
in \fermioncurrents \  
satisfy the following anticommutation relations:
\eqn\fermioncurrentscom{
\{
{\cal J}_{s_1}^a,
{\cal J}_{s_2}^b
\}
= s_1 \epsilon^{ab} \delta_{s_1+s_2,0},
\qquad (a,b= \pm ). 
}
Equation \fermioncurrentscom \ 
remains  also valid for general moding,
using modes 
$s\in Z +1/2 \pm \lambda'$ for $\Phi^{\pm}$,
and so does 
\fermioncurrents \ 
as long as $-1/2 < \lambda' < 1/2$.
This corresponds to twisted b.c.'s
$$
\Phi^{\pm}(\tau,x+l) =
(-1) e^{\mp  2\pi i \lambda'}
\Phi^{\pm}(\tau,x)
$$
The theory is quantized on a vacuum $|0>$ satisfying
\eqn\symplvacuum{
{\cal J}^a_s |0> =0, \quad {\rm for} \quad s\geq 0 . 
}
The character for the general twisted
case reads \kausch
\eqn\chsymplphase{
\chi_{c=-2}(q,\mu,\lambda')
=\Tr~e^{-\beta H}=
q^{-(-2/24)}
q^{-  ({{1\over 2}- \lambda'})^2/2}
\prod_{n=0}^{\infty}
(1+ e^{2 \pi i \mu}
 q^{n+1/2-\lambda' })
(1+  e^{-2 \pi i \mu}
 q^{n+1/2+\lambda' })
.}
As we move
 $\lambda'$ from $0$ to
 $\lambda' =1/2$,
 an extra fermionic zero mode occurs,
whose contribution to the partition function is
analogous to the one for the Dirac fermion.

\vskip0.1in
We see from the expressions 
\betagammaAAla,  \ \complexSAAla \ and \diracphase, \  \chsymplphase \ 
that the bosonic and fermionic characters
 for the $c<0$ and $c>0$ theories  
are identical, respectively, up to multiplicative power
of $q$.
This is no accident. In fact,   
in the following section, we show that the identity
of the characters  is related
to an identity of the space of states of the respective
theories. There is
a `canonical transformation' relating
the 
respective fermionic and bosonic operators,
of the $c>0$ and $c<0$ theories,
in momentum (mode) space.

In Appendix A we summarize the respective characters,
described above,
for the special cases of half-integer  ($\lambda'=0$) and integer
($\lambda'=1/2$) moding.
For these
cases the $c<0$ and $c>0$ fermionic and bosonic characters are completely
identical, respectively \motivBosonZero, including
the multiplicative power of $q$.

\newsec{Mapping between theories with $c>0$ and $c<0$:}
 
The fact that the characters of the $\bgamma$ system and the 
symplectic fermion system 
are identical to those associated with  
the complex scalar and the Dirac 
fermions,  respectively,
in both the half-integer and the integer moded
sectors suggests the existence of
some  `canonical mapping' between the two theories 
at the level of the 
operators that create the space of states,  
used when performing the trace 
giving rise to the respective characters,
$\Tr q^{L_0-c/24}$ ($H = {2 \pi \over l} (L_0-c/24)$ 
is the hamiltonian.).

That is, 
we have  $L_0-c/24$ and $L_0'-c'/24$, 
for the $c>0$ theory and the $c'<0$ theory, respectively,
 in a sector with the same b.c. .
And there is a set of eigenstates for both of these. The characters 
are the same, when we form  the 
trace of $q^{L_0-c/24}$ over 
one set of eigenvectors and the trace of $q^{L_0'-c'/24}$ over the other set 
of eigenvectors. Since the
 trace is invariant under basis change,
one may  expect a 
  similarity transformation between the two sets of 
eigenstates. Thus one would expect a `canonical mapping'
between the
operators 
of the $c>0$ theory and those of the  $c'<0$ theory.

We now construct explicitly this canonical mapping relating the $c>0$ and 
$c<0$ theories. 
It maps the  Hamiltonians  $H = (2 \pi/l)[L_0-c/24]$
into each other. The other   Virasoro generators
$L_n$ for the two theories are 
not simply related,   and the space of
fields of the two theories is different.

\subsec{Bosonic theories: Canonical mapping
between a complex scalar field and bosonic ghost}
\newcount\subsubsectno

We use the notations of section (2).
The bosonic ghost system may be viewed
as arising from an attempt to  construct
a theory of conformal fields in position space,
whose Fourier modes are precisely the
oscillator modes  of the complex scalar
field themselves  (without the
square-roots in \complscalarmodes, \complscalarmodesstar).
This is not possible.  It becomes however possible,
when the following  canonical transformation is performed
on the oscillator modes
of the complex scalar
[$ s\in Z +1/2-\lambda'> 0,$
$  \ \ t\in Z +1/2 +\lambda' >0$]:

$$ B_{+,s}^{\dagger}  \equiv
 {\bar \beta}_s \ \  (= \sqrt{s} \ 
\phi_s^{\dagger} )$$
$$
 B_{+,s}  \equiv { \beta}_s 
 \ \ (= \sqrt{s} \ \phi_s )
$$
$$
 B_{-,t}^{\dagger}  \equiv { \beta}_{-t}
 \ \ (= \sqrt{t} \ \phi_{-t})
$$
\eqn\ghostcanon{
 B_{-,t}   \equiv - {\bar \beta}_{-t}  
 \ \ (= \sqrt{t} \ \phi_{-t}^{\dagger} )
}
where $\phi_s $($s\in Z + 1/2 - \lambda'$) are the Fourier
modes of the complex scalar.
The so-defined  $\beta, \bar \beta$ system is nothing but the bosonic ghost 
theory. Note a crucial minus sign
in the last  equation of \ghostcanon,
which ensures canonical commutation
relations for the $\beta$-modes:
 
$$
[ \beta_{s_2}, {\bar \beta}_{s_1} ]=
\delta_{s_1, s_2}, \quad {\rm for ~all~} 
s_1, s_2 \in Z + 1/2 -\lambda' \quad ({\rm positive \ and \ negative})   
$$
The ``ghost vacuum'', derived this way,   satisfies,
$$
\beta_s  |0> = 0, \quad \bar \beta_{-s}  |0> =0, \quad 
\ s > 0.
$$
( This is  analogous to 
$
\phi_s|0> $ $=\phi_{-t}^{\dagger}|0> =0,$
$ \quad {\rm for} \ \ s>0, t>0 $.)

Using the canonical transformation \ghostcanon \ we see that the 
hamiltonians, $H=(2 \pi/l)[L_0-c/24]$
of the two theories are identical
(giving the results displayed in Section (2)).
 This is easily seen
by  re-writing the complex scalar hamiltonian,
of section (2), in
terms of the bosonic ghost modes, using \ghostcanon
$$
 H^{(\phi,\phi^{\dagger})}/(2 \pi /l)
=
 \sum_{ s \in  Z +{1\over 2} -\lambda', s>0}
 \
s  \
{\bar \beta}_s
 \beta_s
-
 \sum_{ t \in  Z +{1\over 2} +\lambda', t>0}
t \   \beta_{-t} {\bar \beta}_{-t}
$$
$$
=
 \sum_{ s \in  Z +{1\over 2} -\lambda'}
 \
s  \
:{\bar \beta}_s \beta_s:
+ \delta(\lambda')
$$
where $\delta(\lambda')$ is the  shift in
the ground state energy.
For integer moding it is the ${\bar \beta}_s,\beta_s$ ghost
modes with $s \not = 0$ that map to the complex scalar modes
by this transformation, leaving behind  a zero mode oscillator,
decoupled from the complex scalar field [see also the
characters in section (2)].

\subsec{Fermionic theories: relating symplectic fermions and Dirac fermions}
\newcount\subsubsecno

Let us recall the anticommutation relations of the modes 
of the fermionic 
currents, ${\cal J}^a (z) \equiv i \partial_z {\bf \Phi}^a$ 
$$
\{
{\cal J}_{s_1}^a, 
{\cal J}_{s_2}^b
\}
= s_1 \epsilon^{ab} \delta_{s_1+s_2,0}
= |s_1| \quad {\rm sign}(s_1)~\delta_{s_1+s_2,0}, \qquad (a,b= \pm ).
$$
where 
$ s \in Z + 1/2 \pm \lambda $  are mode indices for ${\cal J}^{\pm}_s$.
Comparison with the canonical  anti-commutation relations of 
the modes of the Dirac fermion,  
\eqn\diraccommutation{
\{ \psi^{\dagger}_{s_1}, \psi_{s_2} \}= \delta_{s_1-s_2,0}}
[where $s_1,s_2 \in Z + 1/2 -\lambda'$]
suggests the following mapping: 
$${\cal J}^+_{-s} {\rm sign}(-s) = {\sqrt{|s|}} 
\psi^{ \dagger}_s $$
\eqn\canmapsympldirac{
{\cal J}^-_s  = 
  {\sqrt{|s|}} \psi_s   
}
or equivalently, since ${{\cal J}^\pm_s \over -s}={\bf \Phi}^\pm_s$, 
$$
{\sqrt{|s|}} {\bf \Phi}^+_{-s} = - \psi^{ \dagger}_s
$$
and 
$$
{\sqrt{|s|}} {\bf \Phi}^-_{s} = - {\rm sign}(s)\psi^{}_s.
$$

The vacuum of the symplectic fermion now satisfies 
(Eq.\symplvacuum)
\eqn\diracvacuum{
\psi_{-s}^\dagger |0> =0, \quad \psi_s |0>=0, \quad {\rm for}~s>0}
which is nothing but the Dirac vacuum.

In analogy to the 
the bosonic case, we can use the transformation
\canmapsympldirac \ to express the hamiltonian of the
symplectic fermion theory, given in section (2),
in terms of the Dirac fermion modes.
Using  \hamsympl,
the former reads
when expressed in terms of the modes ${\cal J}^{\pm}_s$:
$$
 L^{symp}_0
=
{-1 \over 2} \sum_s
:[
{\cal J}^+_s
{\cal J}^-_{-s}
-
{\cal J}^-_{-s}
{\cal J}^+_s
]: 
$$
where
$:...:$ denotes (fermionic) normal ordering on the
vacuum defined in \symplvacuum.
Splitting the sum into $s>0,s<0$
 this is equal,
 upto a constant shift in ground state energy, to:
$$=
\sum_{s}
s : \psi^{\dagger}_s \psi_s:
$$
(fermionic normal ordering:
 $ : \psi^{\dagger}_s \psi_s: = \psi^{\dagger}_s \psi_s
$, for $s\geq 0$, and $=-\psi_s \psi^{\dagger}_s $
for $s <0$).
This establishes the identity of the hamiltonians
of the symplectic fermion and the Dirac theories
[up to constant shifts, displayed in section (2)].
At the same time, this establishes,
as mentioned in section  (2),
that the character of the symplectic fermion theory
is identical to that of the Dirac theory.

\newsec{Hidden non-local symmetries}

The existence of a simple mapping
between the modes of the
$c<0$ theories and the $c>0$ theories, 
has very interesting
consequences. It can be used to
exhibit hidden non-abelian
 symmetries
in the simple Dirac theory,
as well as in the theory of the complex
scalar field. Clearly, these two theories
have no obvious non-abelian symmetries.
The origin of these symmetries is
however easily understood from our mapping.

Consider first the bosonic theories.
The bosonic ghost is known to have an $sl(2)_{-1/2}$
Kac-Moody[KM] current algebra symmetry.
The currents can be written as bilinears
in the ghost modes. For half-integer moding,
we use the canonical map
of the previous section, to express
these generators in terms of modes of
the complex scalar.  This establishes
by construction, that the free  complex
scalar field  
possesses $sl(2)$  KM current algebra
symmetry, for half-integer moding.
Particularly simple expressions
are obtained for the {\it global}
 $sl(2)$ generators $I^a_0$ ($a=\pm,3$), the KM zero mode.
We also  establish the existence of a global 
 $sl(2)$  symmetry for the complex scalar
with integer moding.
Under this $sl(2)$, the complex scalar field
$(\phi,\phi^{\dagger})$ transforms as a doublet.
Interestingly, the $I_0^{\pm}$ generators
are non-local, whereas $I_0^3$ is local,
in the complex scalar field variables.
In the Hilbert space of the complex
scalar,
the $sl(2)$ symmetry 
rotates
the two sets of oscillator modes
of the complex scalar into each other.
In addition we observe, interestingly, that
the  Hilbert space has the same
structure as that encountered
in the Yangian description of the level-1 $sl(2)$
current algebra \refs{\haldane, \schout, \pasquier}. This suggests the 
existence of a Yangian symmetry algebra
for both, the complex scalar and the  bosonic ghost
theory.

\nref\leclair{A. Leclair, Nucl. Phys. B415 (1994) 734.} %
As discussed above, the fermionic $c=-2$  theory 
possess a global symplectic $Sp(2)$  symmetry \refs{\readmilan,\kausch}.
This does not extend to a current algebra symmetry
(at least not in an obvious way,
since the corresponding Noether current has
logarithmic factors in its correlation functions).
 Mapping
 the corresponding symmetry generators
into the Dirac fermion language, we
establish that the latter also 
 has  a nonabelian  (global) $sl(2)$
symmetry. The pair $(\psi^{\dagger},\psi)$ transforms
as a doublet under this  symmetry.
Again, the corresponding $Q^{\pm}$
generators are non-local, whereas
$Q^3$ is the ordinary local Dirac charge
\commentLeClair.

We now discuss bosonic and fermionic non-abelian symmetries
in turn.

\subsec{Complex scalar theory}
\newcount\subsubsecno
The action of the complex scalar field \scalaraction \ 
has an obvious $U(1)$ symmetry.
However, the associated Noether current (we consider again
only the L-chiral part)
$ J(\tau,x) =
 {1\over 2 } [ \phi^*\partial_z \phi
-
 \phi\partial_z \phi^*]
$
does not generate a current algebra,
due to logarithmic factors in its correlation functions.
Nevertheless, of course, its spatial integral
is the conserved $U(1)$ charge
\eqn\uONEglobal{
J_0 = {1\over 2 i}
\int d x 
 [ \phi^{\dagger}\partial_x \phi
-
 \phi\partial_x \phi^{\dagger}]
}

We now show that, for half-integer moding,
 the global 
$U(1)$ generator
\uONEglobal \ is the zero mode of 
a $U(1)$ KM current algebra, using the mapping
from the $c=-1$ ghost theory. Moreover, this procedure
will give us  a set of
$sl(2)$  current algebra generators of the complex scalar theory,
for half-integer moding. 
For integer moding, we will construct  {\it global}
 $sl(2)$ generators for the complex scalar.

The $c=-1$ bosonic ghost theory possesses 
three conformal weight $=1$ currents:
$$
I^+(z) \equiv {1\over 2} {\bar \beta}^2(z),
\quad
I^-(z) \equiv {-1\over 2} {\beta}^2(z),
\quad
I^3(z) \equiv {1\over 2} :{\bar \beta}{\beta}:(z).
$$
Using the correlator of the ghost fields, 
$<{\bar \beta}(z_1)\beta(z_2)> =$
$ -< \beta(z_1) {\bar \beta}(z_2) > =$ $ 1/(z_1-z_2)$ 
one finds for the singular part of  the short distance expansions
$$
I^{\pm}(z)
I^{\mp}(0) \sim  {k \over z^2} + { \pm 2 I^3(0)\over z}
$$
$$
I^{\pm}(z)
I^3(0) \sim { \pm I^{\pm} \over z} 
$$
$$
I^3(z)I^3(0)\sim {k/2 \over z^2} 
$$ with
$k=-1/2$.
Hence, the modes
$$
I_n^+
=
{1\over 2}
\sum_s {\bar \beta}_s{\bar \beta}_{-s-n}
$$
$$
I_n^-
=
{-1\over 2}
\sum_s {\beta}_{s+n}{\beta}_{-s}
$$
\eqn\suTWOghostKM{
I_n^3
=
{1\over 2}
\sum_s :{\bar \beta}_s{\beta}_{s+n}:
}
of the currents 
satisfy an ${sl(2)}$
Kac-Moody current algebra
at level $k=-1/2$.

We focus first on half-integer moding.
Expressing these modes in terms of  those
of the complex scalar, using \ghostcanon,
gives  us
a set of $sl(2)$ current algebra
generators that commute with the hamiltonian of the
complex scalar field in \scalarhamiltonian.
The presence of a current algebra symmetry
of the complex scalar field seems rather unexpected.

Let us now focus in more detail on the global
$sl(2)$ generators, induced by this mapping on
the scalar field theory.
Using this mapping, 
we express
the global generators $I^a_0$ ($a=\pm,3$)
of \suTWOghostKM \ 
 in terms of the creation/annihilation operators,
of the complex scalar.
First consider the case of half-integer moding,
for which we find
$$
I_0^+=(-1) \sum_{s>0} B_{+,s}^{\dagger} B_{-, s}
$$
$$
I_0^-=(-1) \sum_{s>0}  B_{-, s}^{\dagger} B_{+, s}
$$
\eqn\suTWOoscillatorglobal{
I^3_0
=
{1\over 2}
\sum_{s>0}
[
B_{+,s}^{\dagger} B_{+, s}
-
B_{-,s}^{\dagger} B_{-, s}
]
}
[where $s \in Z + 1/2$].
We see that  the global  $sl(2)$ symmetry
rotates
the two sets of oscillator  modes
of the complex scalar field into each other.
It is this symmetry that  extends to the $sl(2)$
current algebra symmetry.
The expressions for the modes
$I^a_n$ with $n \not = 0$
 are not quite  so simply expressed in
terms of the oscillator modes; nevertheless
they form an $sl(2)$ current algebra, by construction.

We observe that the Hilbert space possesses
 a structure isomorphic
to the  multiplets of
 the (nonlocal)  Yangian symmetry
present in the $sl(2)$ current algebra
representations {\it at level $k=1$} 
\refs{\haldane, \schout, \pasquier}: 
Consider a general state
in the complex boson Hilbert space 
\eqn\yangianmotifs{
B^{\dagger}_{\alpha_N, {1\over 2}+n_N}
B^{\dagger}_{\alpha_2, {1\over 2}+n_2}
B^{\dagger}_{\alpha_1, {1\over 2}+n_1}|0>,
\qquad
n_N\geq ...
n_2\geq ...
n_1 \geq 0
}
where $\alpha_i=\pm$ ($i=1,...,N$).
This state has $L_0$-eigenvalue
$N/2 +\sum_{i=1}^N n_i$.
Consider a fixed set of mode indices $n_i$.
When all $n_i$
 are different
this is a $2^N$ dimensional space.
On the other hand,
if some mode indices are equal, the number of states
is reduced, the corresponding product of  $sl(2)$
doublets with equal mode index  being projected on the totally symmetric
combination only (due to bose statistics).
For example, for $N=2$, there are four states
when $n_1\not = n_2$,
but only three states when $n_1=n_2$.
This is the same structure of Hilbert space \schoutthanks \ 
encountered in the Yangian description of the
$sl(2)_1$ current algebra modules \refs{\haldane,\schout, \pasquier},
suggesting the existence of Yangian
symmetry generators in the complex scalar
and bosonic ghost theory.
Interestingly, this structure is realized in the `permanent'
Quantum Hall state \refs{\readmoore,\readmilan}.

Finally, consider
the expressions for
the  global $sl(2)$ generators in terms of the complex
scalar  field itself.
When expressed in terms
of the momentum modes of the scalar field $\phi_s,\phi^{\dagger}_s$,
we get
$$
I^+_0
=
{-1\over 2}
\sum_s |s| \phi^{\dagger}_s\phi^{\dagger}_{-s}
$$
$$
I^-_0
=
{-1\over 2}
\sum_s |s| \phi_s\phi_{-s}
$$
\eqn\complexscalarSLtwo{
I^3_0
=
{1\over 2}
\sum_s s :\phi^{\dagger}_s \phi_s:
}
We see that whereas 
$I^3_0$, which
is just $(1/2)$  times the charge generator
\uONEglobal, can be locally expressed in terms of the 
the  complex scalar field,
the other generators $I^{\pm}_0$
give rise to non-local expressions in the   complex
scalar  field variables in position space.

For integer moding
the generators $I^a_n$
 of the bosonic ghost
theory
[Eq.\suTWOghostKM]
 still generate 
 a current algebra (see Appendix B for a brief
discussion).
Let us focus on the {\it global} generators
$I^a_0$.
These can be split into a sum
of  two commuting sets of  generators,
a ghost zero mode contribution $j^a$, and
a piece  ${\tilde I}^a$
that comes from all $s \not = 0$ modes,
$ I^a_0 = j^a +{\tilde I}^a$.
 The generators ${\tilde I}^a$
can be mapped into the complex scalar theory,
using \ghostcanon. They give rise to a global $sl(2)$
symmetry, acting on the momentum modes
with non-zero mode index
(which still has the structure
discussed below  \yangianmotifs). 
These can be expressed in terms of the
complex scalar field itself as in\complexscalarSLtwo.
Again, the last generator in that equation is
the ordinary $U(1)$ charge of the complex  scalar.
In summary, we established that the complex scalar field (at zero boson
radius) has a global $sl(2)$ symmetry, for integer moding.

\subsec{Dirac fermions}
\newcount\subsubsecno

The symplectic fermion action is invariant under $sl(2)$, under which the 
$\Phi^\pm$ transform as a doublet. The Noether currents associated with 
this symmetry are \refs{\readmilan,\kausch}  
\eqn\sltwocurrent{
J_L^\alpha(z, z^*)={i \over 2} t^\alpha_{ab} : \Phi_L^a \partial_z \Phi_L^b
}
and likewise for $J_R$. We treat only the $L$-moving
(holomorphic) sector again, dropping the 
subscript. 

Following the notation of \kausch ,
we chose the following basis
 for the  $sl(2)$ generators, 
$t^\alpha_{ab}$:
\eqn\sltwobasis{ 
t^0_{ab}= {1 \over 2}\pmatrix{ -1& 0 \cr
0 & -1 \cr}, \quad 
t^1_{ab}={1 \over 2}\pmatrix{ 1& 0 \cr
0 & -1 \cr}, \quad
t^2_{ab}={1 \over 2}\pmatrix{ 0& -1 \cr
-1 & 0 \cr}.}

The $sl(2) $ symmetry is  present
for antiperiodic ($\lambda'=0$)
and periodic ($\lambda'=1/2$) b.c.'s.
In the following expressions, the sums 
run either over $s \in Z + 1/2$, or over $s \in Z, s \not = 0$. 
The $sl(2)$ charges $Q^\alpha$ can be computed explicitly \kausch \ 
and we write 
them in terms of 
$\Phi^\pm$,  or  the fermionic currents ${\cal J}^\pm$. 
\eqn\sltwocharges{
Q^\alpha=t^\alpha_{ab} \sum_{s} {{\cal J}^a_{-s} {\cal J}^b_s \over s}
}
where $\alpha =0,1,2$ and $a,b=\pm$. 
The $sl(2)$ charges of the $\Phi^\pm$ theory
 can now be 
translated into corresponding charges in the Dirac theory using the 
canonical mapping between the fermions $\Phi^\pm$ and Dirac 
fermions.
Explicitly, we have:
$$
Q^0=t^0_{++} \sum_{s}  {{\cal J}^+_{s} 
{\cal J}^+_{-s} \over -s}
+ t^0_{--} \sum_{s}  
{{\cal J}^-_{-s} 
{\cal J}^-_{s}\over s}
$$
with $t^0_{++}=t^0_{--}=-1/2$. The charge $Q^0$ in terms of the Dirac 
fields is {\it non-local}: 
$$  
Q^0={-1/2} \sum_{s} {\rm sign}(s) 
\Bigl [ 
\psi^{ \dagger}_{-s} \psi^{ \dagger}_{s} 
+
 \psi_{-s} \psi_{s}
\Bigr ]. 
$$
The charge $Q^1$ is given by 
$$
Q^1=t^1_{++} \sum_{s} {{\cal J}^+_{s}  
{\cal J}^+_{-s} \over -s}
- t^1_{--} \sum_{s} {{\cal J}^-_{-s}  
{\cal J}^-_{s}\over s}. 
$$
with $t^1_{++}= - t^1_{--}=1/2$ and in the Dirac fermion language, this 
becomes
$$ 
Q^1=
{1/2}
 \sum_{s} 
{\rm sign}(s) 
\Bigl [ 
\psi^{ \dagger}_{-s} \psi^{ \dagger}_{s} 
-
\psi_{-s} \psi_{s}
\Bigr ] 
$$
which is again {\it non-local}. 
Finally,  
$$
Q^2=t^2_{+-} \sum_{s} {{\cal J}^+_{s}  
{\cal J}^-_{-s} \over -s}
+ t^2_{-+} \sum_{s} {{\cal J}^-_{-s}  
{\cal J}^+_{s}\over s}
$$
with $t^2_{+-}=t^2_{-+}=-1/2$, giving, in terms of Dirac fermions, 
a {\it local} charge: 
$$ 
Q^2={-1/2} \sum_{s}\Bigl [ 
-\psi^{ \dagger}_{-s} 
 \psi_{-s}
+\psi_{-s} 
\psi^{ \dagger}_{-s} 
 \Bigr ]
=\sum_{s} [ \psi^{\dagger}_s \psi_s -{1\over 2} ].
$$

In {\it summary}, 
using the basis $Q^{\pm},Q^3$,
 the global $sl(2)$ generators of the 
symplectic fermionic 
system have the following form in the theory of free Dirac fermions: 
$$
 Q^+ = \sum_{s>0} \psi^{\dagger}_{-s}\psi^{\dagger}_{s}
$$

$$
 Q^- = \sum_{s>0} \psi_{s}\psi_{-s}
$$

\eqn\fermionicSUtwo{ 
Q^3=
{1\over 2} \sum_{s} [ \psi^{\dagger}_s \psi_s -{1\over 2} ].
}

$$
\bigl ( [Q^+,Q^-]= 2 Q^3, \quad  [Q^3, Q^{\pm}] = \pm Q^{\pm}
\bigr )
$$
[As mentioned above, the last sum
excludes the zero mode, $s=0$, for integer moding.]

\vskip0.1in
We end this section with a simple application of the non-local 
symmetries: Consider a Dirac fermion on a half infinite space,
initially with a boundary condition
$\psi_L(x=0) =\psi_R(x=0)$.
We wish to add a term
to the action  that  would give a
boundary condition
$\psi^{\dagger}_L(x=0) =C\psi_R(x=0) +S \psi^{\dagger}_R(x=0) 
$, $|C|^2 + |S|^2=1$. This is  achieved
by addition of  a non-local  term

$$\delta {\cal A} =
(\lambda/2i) \int d y_1\int d y_2
 {1 \over (y_1-y_2)}
[\psi^{\dagger}(x=0,y_1)\psi^{\dagger}(x=0,y_2)
-
\psi(x=0,y_1)\psi(x=0,y_2)]
$$
(where $y$ is the coordinate along the boundary),
corresponding to a linear combination of the
two non-local generators.

\newsec{Conclusions and comments}

In this paper we have found a canonical mapping
 between the $c=-1$ bosonic ghost theory and the $c=+2$ complex scalar 
theory and also between the $c=-2$ theory of ``symplectic'' fermions and 
the $c=+1$ Dirac fermion theory, mapping the
Hamiltonians
$H=(2 \pi/l)[L_0-c/24]$ into each other.
 As a consequence, the 
characters of the respective theories that are related by this 
canonical 
mapping are the same in a general ``twisted'' sector upto a power of 
$q=e^{-2\pi \beta/l}$ 
and are identical in the sectors with twist $\lambda'=0, 1/2$
(antiperiodic, and periodic). Moreover, 
this canonical mapping reveals a 
{\it hidden   $sl(2)$ current algebra symmetry}  of 
the complex scalar  
for half-integer moding,
as well as a
{\it hidden  global  $sl(2)$ symmetry} 
for integer moding and
 for the Dirac fermion theories, that these theories 
inherit from their $c<0$ counterparts. One of
the global generators 
in these two theories is the  $U(1)$ charge.
The remaining generators
are {\it non-local} when expressed
 in the language of the $c>0$ fields and 
are therefore not  visible from the Lagrangian
(i.e.  without the knowledge
 of the mapping with the $c<0$ theories that
we use).
\vskip0.1in
The fact that the non-unitary $c<0$ theories have the
same space of states  and the same hamiltonian
as their unitary counterparts, permits 
the $c<0$ ghost theories to represent
physical paired Quantum Hall states \refs{\readmilan,\readnew}, which
require a positive definite Hilbert space.
It is remarkable that, for these systems,
the  {\it bulk} Quantum Hall wave functions are represented
by conformal blocks of correlation functions
of the $c<0$ theories \readmoore,
 whereas the hamiltonian 
of the quantum hall  {\it edge states }
associated with
these bulk wavefunctions possesses the
Hilbert space of the {\it unitary} $c=1$ and $c=2$ 
theories.  This is clearly necessary for those
theories to represent {\it physical}
Quantum Hall systems. The correlation functions
of the edge mode fields are also  those
of a $c<0$ theory. It would be interesting
to study further examples of conformal field theories,
 that
{\it are } non-unitary, but, at the same time,
share their space of states,
and Hamiltonian  with a  {\it unitary} theory.

The presence of non-abelian symmetries, in the $c=1,2$ theories
is unexpected.  The non-locality of the symmetry generators
has the `flavor' of Yangian symmetries, whose generators
are generally non-local. Yangian symmetries
are thought to be responsible for the
the integrability of Bethe Ansatz solvable models.
[In a sense, they are the ``symmetries''
of  Bethe Ansatz integrability].
  At a more general level,
there has been much discussion recently,
about  different bases for a given conformal field theory.
Some physical applications have come from
viewing a certain conformal field theory as a 
Yang-Baxter (Bethe Ansatz) integrable system.
Integrability, then, generates
a basis of the Hilbert space of the conformal field
theory. Such bases are in general not the Verma
module bases, but of a very different kind, in that they
possess   a 
Fock-space like structure. The Verma module basis
is isomorphic to the space of fields.
The Bethe Ansatz basis  is a basis for the
space of states. In general,   these bases are completely different in
structure. 
The non-unitary ghost theories
provide thus examples of a similar phenomenon.
The possibility to
view a given conformal field theory in
different bases, has been at the root of much
progress, in solving strong coupling
problems in interacting theories.
The reason is simply that the interaction
may look simple in a suitably chosen basis,
and become analytically tractable. 
One set of such  examples are various
Kondo models \AffLudwig. Another example is recent
progress on Quantum Hall point contacts
devices \FLS, involving the Bethe Ansatz basis mentioned.
\vskip0.1in
 We hope that our analysis of the
ghost theories will lend some intuition
that would help us get a better understanding
of random systems in the strong coupling regime.
Some aspects of this will be addressed
 in a separate piece of work \GLrandom.

\vskip0.1in

\noindent{\bf Acknowledgements} A.W.W.L thanks the ICTP, Trieste (Italy)
for their kind hospitality, and acknowledges support
as a Fellow from the
A. P. Sloan  Foundation.

\appendix{A}{Characters for half-integer and integer moding}

In this appendix we summarize the respective characters
for the cases of half-integer  ($\lambda'=0$) and integer
($\lambda'=1/2$) moding. 
As we mentioned in section 2, for these 
cases, the $c=-1$ character of the $\bgamma$ system and the $c=2$ 
character of the complex scalar theory are identical \motivBosonZero.
Similarly, so are the $c=-2$ character of the ``symplectic fermions'' and 
the $c=1$ character of the Dirac fermions.  

We 
denote the characters by the boundary conditions in time and space,
$(B_{\tau},B_{x})$, when $\mu=0$.
The variable $\mu$ then defines a non-specialized
character, permitting us to extract more detailed
information about the space of states.

\vskip0.1in 
\subsec{Bosonic characters}

\noindent{\bf A.1.1. Half-integer moding($\lambda'=0$)}
\vskip0.1in
\noindent\underbar{\it  Bosonic ghost}

\eqn\betagammaAA{
\chi^{\bgamma,AA}_{c=-1}
(q,\mu)
=
q^{-({-1\over 24} )}
\prod_{n=0}^{\infty}
{1
\over 
(1+ e^{ 2 \pi i \mu} q^{n+1/2})
}
 \ 
{
1 \over (1+ e^{-2\pi i \mu} q^{n+1/2})
},
}
Lowest  $c=-1$ conformal weights: 
$ \Delta^{\bgamma,AA}_{c=-1} = 0, 1/2, 1,...  $.
\vskip0.1in

\noindent \underbar{\it Complex scalar } 
\vskip0.1in

\eqn\CscalarAA{
\chi^{(\phi^*, \phi),AA}_{c=2}(q, \mu)
=
q^{- 2/24}
 \ q^{1/8} \ 
\prod_{n=0}^{\infty}
{1\over (1+ e^{2\pi i \mu} q^{n+1/2})}
 \ 
{1 \over (1+ e^{-2\pi i \mu} q^{n+1/2})},}
Lowest $c=+2$ conformal weights:
$ \Delta^{(\phi^*,\phi), AA}_{c=2} = 1/8, 1/8 +1/2, 1/8 + 1,...  $.
Note that the prefactor  can be written as
\eqn\prefactor{
q^{- 2/24}
 \ q^{1/8}  =
q^{-(-1/24)}
}
exhibiting the identity
of the  $c=-1$ and the $c=+2$
bosonic  characters:
\eqn\ghostequalscalar{
\chi^{(\phi^*, \phi),AA}_{c=2}(q,\mu)
=
\chi^{\bgamma,AA}_{c=-1}(q,\mu).}
We could also use the
partition function of the complex scalar theory 
to cancel the partition function
of the Dirac theory. In the first case,
the combined (fermion plus boson) theory
has $c=3$, and $N=2$ superconformal
symmetry. In the latter case
 the combined theory has $c=0$,
and graded SUSY. (The partition function
is unity, in both cases.)

\vskip .7cm
 
\noindent{\bf A.1.2. Integer moding ($\lambda'=1/2$)}
\vskip0.1in

\noindent\underbar{\it Bosonic ghost}
\vskip .3cm
\eqn\betagammaAP{
\chi^{\bgamma,AP}_{c=-1}(q,\mu)
=
q^{-(-1/24)} \ 
q^{-1/8} \ 
{1\over (1+ e^{2\pi i \mu})}
\prod_{n=1}^{\infty}
{1\over (1+ e^{2\pi i \mu} q^{n})}
 \ 
{1 \over (1+ e^{-2\pi i \mu} q^{n})},}
the first factor being the contribution
from the bosonic zero mode.  Lowest  $c=-1$ conformal weights:
$ \Delta^{\bgamma, AP}_{c=-1} = -1/8, -1/8 +1,...
$.
\vskip .4cm
\noindent\underbar{\it Complex Scalar}
\vskip .3cm
\eqn\CscalarAP{
\chi^{(\phi^*,\phi),AP}_{c=2}(q,\mu)
=
q^{- 2/24}
\ 
{1\over (1+ e^{2\pi i \mu})}
\prod_{n=1}^{\infty}
{1
\over
 (1+ e^{2\pi i \mu} q^{n})
}
 \ 
{1 \over
 (1+ e^{-2\pi i \mu} q^{n} )
},}
Lowest $c=+2$ conformal weights: 
$ \Delta^{(\phi^*, \phi), AP}_{c=2} = 0, 1, 2,...  $.

\vskip .2cm
\noindent Again, we find that the two  bosonic characters
are identical:
$$
\chi^{\bgamma,AP}_{c=-1}(q,\mu)
=
\chi^{(\phi^*,\phi),AP}_{c=2}(q,\mu)
$$

\vskip .5cm

\subsec{Fermionic characters} 

\noindent{\bf A.2.1. Half-integer moding} 

\vskip .2cm

\noindent\underbar{\it Dirac fermion}
\eqn\DiracAA{
\chi^{Dirac,AA}_{c=1} (q,\mu)
=
q^{-1/24} \ 
\prod_{n=0}^{\infty}
(1+ e^{2 \pi i \mu}
 q^{n+1/2})
(1+  e^{-2 \pi i \mu}q^{n+1/2})}
Lowest $c=1$ conformal weights:
$
\Delta^{Dirac,AA}_{c=1} = 0,1/2,...
$.

\vskip .4cm

\noindent\underbar{\it Symplectic Fermion}
\vskip0.1in

\eqn\symplAA{
\chi^{sympl,AA}_{c=-2} (q,\mu)
=
q^{-(-2/24)} \ q^{-1/8}
\prod_{n=0}^{\infty}
(1+ e^{2 \pi i \mu}
 q^{n+1/2})
(1+  e^{-2 \pi i \mu}
q^{n+1/2}),}
Lowest $c=-2$ conformal weights: 
$
\Delta^{sympl,AA}_{c=-2} = -1/8, -1/8 + 1/2,...
$.
Eq.\prefactor \ 
then shows the identity of
this and the Dirac 
character,
\eqn\diracAAequalssympl{
\chi^{Dirac,AA}_{c=1} (q,\mu)
=
\chi^{sympl,AA}_{c=-2} (q,\mu)}

\vskip .5cm

\noindent{\bf A.2.2. Integer moding }
\vskip0.1in

\noindent\underbar{\it Dirac fermion}

\eqn\DiracAP{
\chi^{Dirac,AP}_{c=1} (q,\mu)
=
q^{-1/24} \
q^{1/8}  
\
(1+ e^{2 \pi i \mu})
\prod_{n=1}^{\infty}
(1+ e^{2 \pi i \mu} q^{n})
(1+  e^{-2 \pi i \mu} q^{n}),
}
where the extra factor comes from the zero mode.
Lowest $c=1$ conformal weights:
$
\Delta^{Dirac, AP}_{c=1}
= 1/8, 1/8  +1,...$.

\vskip .5cm

\noindent\underbar{\it Symplectic Fermion}

\eqn\symplAP{
\chi^{sympl,AP}_{c=-2} (q,\mu)
=
q^{-(-2/24)}
(1+  e^{2 \pi i \mu})
\prod_{n=1}^{\infty}
(1+ e^{2 \pi i \mu}
 q^{n})
(1+  e^{-2 \pi i \mu}
q^{n}),}
Lowest $c=-2$ conformal weights:
$
\Delta^{sympl, AP}_{c=-2}
=0, 1,2,...
$.
Again Eq.\prefactor \ show the exact identity of
this and the Dirac character:
\eqn\DiracAPequalssympl{
\chi^{Dirac,AP}_{c=1} (q,\mu)
=
 \ \chi^{sympl,AP}_{c=-2} (q,\mu)}

\vskip0.2in

\appendix{B}{ Characters of the
$sl(2)_{-1/2}$ current algebra 
}

\vskip .3cm

In this Appendix  we  briefly discuss 
the relationship of the 
characters of the bosonic ghost
theory of Appendix A, for both integer and half-integer moding,
with the expressions
obtained by Kac and Wakimoto  for the
fractional level 
$sl(2)_{-1/2}$ current algebra
obeyed by the generators $I^a_n$ (Eq.\suTWOghostKM). 
Projecting the
characters (A.1) of Appendix A,
onto even and odd ghost boson number,
we obtain two characters 
at $c=-1$ with scaling dimensions
$\Delta=0,1/2$ (modulo integers):
$$
\chi^{\bgamma, 0}_{c=-1}\equiv
{1\over 2} 
[ \chi^{\bgamma,AA}_{c=-1} (q,\mu)
+
\chi^{\bgamma,AA}_{c=-1} (q,\mu+1/2)
]
$$
$$
\chi^{\bgamma, 1/2}_{c=-1}\equiv
{1\over 2} 
[ \chi^{\bgamma,AA}_{c=-1} (q,\mu)
-
\chi^{\bgamma,AA}_{c=-1} (q,\mu+1/2)
]
$$
Similarly those with periodic spatial b.c.'s,
give two characters  
at $c=-1$  with scaling dimensions
$\Delta=-1/8$ (modulo integers):
$$
\chi^{\bgamma, -1/8,+}_{c=-1}\equiv
{1\over 2} 
[ \chi^{\bgamma,AP}_{c=-1} (q,\mu)
+
\chi^{\bgamma,AP}_{c=-1} (q,\mu+1/2)
]
$$
\eqn\projections{
\chi^{\bgamma, 1/8,-}_{c=-1}\equiv
{1\over 2} 
[ \chi^{\bgamma,AP}_{c=-1} (q,\mu)
-
\chi^{\bgamma,AP}_{c=-1} (q,\mu+1/2)
]
}

On the other hand, the characters of the  fractional
level $sl(2)_{-1/2}$ current algebra have been
derived by Kac and Wakimoto \kacwakimoto,
and expressed in terms of theta functions.
There are four characters, $\chi^{KW}_{n,k}(q,\mu)$
with $n,k=0,1$, transforming into each other
under modular transformations.
Comparison of those with the expressions
in \projections \ shows that 
$\chi^{KW}_{0,0}$,
$\chi^{KW}_{1,0}$,
$\chi^{KW}_{0,1}$,
$\chi^{KW}_{1,1}$
correspond precisely  to 
$\chi^{\bgamma, 0}_{c=-1}$,
$\chi^{\bgamma, 1/2}_{c=-1}$,
$\chi^{\bgamma, 1/8, +}_{c=-1}$,
$\chi^{\bgamma, 1/8, -}_{c=-1}$,
respectively.

{}From the comparison with the characters
in \projections, the interpretation of the
Kac-Wakimoto characters is as follows:
For half-integer moding,
the generators \suTWOghostKM \ 
act on the Hilbert space without zero mode.
The even- and odd- boson  number characters
correspond thus  to 
$\chi^{KW}_{0,0}$,
$\chi^{KW}_{1,0}$. 
 On the eigenspace of lowest $L_0$ eigenvalue
the global generators $I^a_0$
transform in a unitary representation of $SU(2)$
 of spin $=0$
and $=1/2$ respectively, for even and odd boson numbers.
These characters correspond therefore
to $SU(2)_{-1/2}$ primaries.

For integer moding, on the other hand,
the generators \suTWOghostKM \ 
act on the Hilbert space  with the extra zero mode boson
$\beta_0|0>=0$. The  eigenspace of lowest
$L_0$ eigenvalue is now infinite dimensional,
corresponding to the states ${\bar \beta}_0^{N}|0>$, (N=0,1,2,...).
When splitting into even and odd $N$,
these are two   unitary irreducible lowest weight
representations of $SU(1,1)$. 
The global generators $I^a_0$ do not act unitarily
on this space. However, the generators
 re-defined  by 
$ I^+_n \to {\hat I}^+_n$, $ I^-_n \to (-1) {\hat I}^-_n$,
$I^3_n \to  {\hat I}^3_n$,
satisfy an $SU(1,1)$ current algebra \commentreadsuOneOne. In particular,
the global generators
${\hat I}^a_0$
act unitarily on the space of even and odd boson zero modes,
and correspond to the (infinite dimensional)
irreducible  lowest weight representions
of ``angular momentum'' $j=-3/4$ and $j=-1/4$ \suOneOne.
The (integer moded) characters
$\chi^{KW}_{0,1}$,
$\chi^{KW}_{1,1}$ 
correspond to these representations
of the $SU(1,1)_{1/2}$ current algebra.

\listrefs

\end